\documentclass[aps,prd,10pt,twocolumn, superscriptaddress]{revtex4}
\usepackage[dvips]{graphicx}
\usepackage{epsfig, comment, caption}
\usepackage{multirow}
\usepackage{latexsym,amsmath,amsfonts,amssymb}
\usepackage{mathtools} 
\usepackage{slashed} 
\usepackage{blkarray}
\usepackage[all]{xy} 
\usepackage[makeroom]{cancel}
\usepackage[latin1]{inputenc}
\usepackage[american]{babel}
\usepackage{bbm}
\usepackage{color}
\usepackage[unicode]{hyperref}
\usepackage{soul}
\usepackage{ulem}

\hypersetup{ linktoc=all,
    colorlinks, linkcolor={palatinateblue},
    citecolor={brightpink}, urlcolor={amaranth}
}

\graphicspath{{Images/}}

\definecolor{rosy}{RGB}{230,235,252}
\definecolor{myframetitle}{RGB}{90,89,170}
\definecolor{myblocktitle}{RGB}{140,185,249}
\definecolor{mytitle}{RGB}{10,80,26}

\definecolor{darkgreen}{RGB}{27,130,45}
\definecolor{darkblue}{rgb}{0,0,0.3}
\definecolor{darkred}{rgb}{0.7,0,0}

\definecolor{light gray}{RGB}{220,220,220}
\definecolor{dark purple}{RGB}{108,0,217}
\definecolor{pink}{RGB}{190,20,100}
\definecolor{orang}{RGB}{193,63,0}
\definecolor{green}{RGB}{11,98,17}
\definecolor{darkpink}{RGB}{153,0,76}
\definecolor{bluegreen}{RGB}{0,102,102}
\definecolor{greenlagan}{RGB}{0,102,0}
\definecolor{redgreen}{RGB}{102,102,0}
\definecolor{Redgreen}{RGB}{153,76,0}
\definecolor{vividviolet}{rgb}{0.62, 0.0, 1.0}
\definecolor{amaranth}{rgb}{0.9, 0.17, 0.31}
\definecolor{palatinateblue}{rgb}{0.15, 0.23, 0.89}
\definecolor{brightpink}{rgb}{1.0, 0.0, 0.5}
\definecolor{cornflowerblue}{rgb}{0.39, 0.58, 0.93}
\definecolor{deepcarminepink}{rgb}{0.94, 0.19, 0.22}
\definecolor{radicalred}{rgb}{1.0, 0.21, 0.37}

\def\6{\partial}

\newcommand{\be}{\begin{equation}}
\newcommand{\ee}{\end{equation}}
\newcommand{\beq}{\begin{equation}}
\newcommand{\eeq}{\end{equation}}
\newcommand{\bea}{\begin{eqnarray}}
\newcommand{\eea}{\end{eqnarray}}

\newcommand{\ba}{\begin{eqnarray}}
\newcommand{\ea}{\end{eqnarray}}

\newcommand{\beqs}{\begin{eqnarray}}
\newcommand{\eeqs}{\end{eqnarray}}
\newcommand{\bal}{\begin{aligned}}
\newcommand{\eal}{\end{aligned}}

\def\lbldef#1#2{\expandafter\gdef\csname #1\endcsname {#2}}

\def\href#1#2{#2}

\newcommand{\ber}{\begin{eqnarray}}
\newcommand{\eer}{\end{eqnarray}}

\newcommand{\beqar}{\begin{eqnarray}}

\newcommand{\eeqar}{\end{eqnarray}}

\newcommand{\dsl}
   {\kern.06em\hbox{\raise.15ex\hbox{$/$}\kern-.56em\hbox{$\partial$}}}

\newcommand{\eeqarr}{\end{eqnarRAy}}
\newcommand{\ZZ}{{\rm \kern 0.275em Z \kern -0.92em Z}\;}

\def\CC{{\mathchoice
{\rm C\mkern-8mu\vrule height1.45ex depth-.05ex
width.05em\mkern9mu\kern-.05em}
{\rm C\mkern-8mu\vrule height1.45ex depth-.05ex
width.05em\mkern9mu\kern-.05em}
{\rm C\mkern-8mu\vrule height1ex depth-.07ex
width.035em\mkern9mu\kern-.035em}
{\rm C\mkern-8mu\vrule height.65ex depth-.1ex
width.025em\mkern8mu\kern-.025em}}}

\def\RR{{\rm I\kern-1.6pt {\rm R}}}

\def\ZZ{{\rm Z}\kern-3.8pt {\rm Z} \kern2pt}
\def\IB{\relax{\rm I\kern-.18em B}}
\def\ID{\relax{\rm I\kern-.18em D}}
\def\II{\relax{\rm I\kern-.18em I}}
\def\IP{\relax{\rm I\kern-.18em P}}

\newcommand{\bear}{\begin{eqnarray}}
\newcommand{\eear}{\end{eqnarray}}

\def\6{\partial}

\newfont{\namefont}{cmr10}
\newfont{\addfont}{cmti7 scaled 1440}
\newfont{\boldmathfont}{cmbx10}
\newfont{\headfontb}{cmbx10 scaled 1728}

\allowdisplaybreaks[1]

\numberwithin{equation}{section}

\begin{document}

\title{Hints of FLRW Breakdown from Supernovae}

\author{Chethan Krishnan} \email{chethan.krishnan@gmail.com}
\affiliation{Center for High Energy Physics, Indian Institute of Science, Bangalore 560012, India}
\author{Roya Mohayaee} \email{mohayaee@iap.fr}
\affiliation{Sorbonne Universit\'e, CNRS, Institut d'Astrophysique de Paris, 98bis Bld Arago, Paris 75014, France}
 \author{Eoin \'O Colg\'ain}\email{ocolgain@gmail.com}
 \affiliation{Center for Quantum Spacetime, Sogang University, Seoul 121-742, Korea}
 \affiliation{Department of Physics, Sogang University, Seoul 121-742, Korea} 
 \author{M. M. Sheikh-Jabbari}\email{shahin.s.jabbari@gmail.com}
\affiliation{School of Physics, Institute for Research in Fundamental Sciences (IPM), P.O.Box 19395-5531, Tehran, Iran}
\author{Lu Yin}\email{yinlu@sogang.ac.kr}
\affiliation{Center for Quantum Spacetime, Sogang University, Seoul 121-742, Korea}
 \affiliation{Department of Physics, Sogang University, Seoul 121-742, Korea}

\begin{abstract}
A 10\% difference in the scale for the Hubble parameter constitutes a clear problem for cosmology. Here, considering angular distribution of Type Ia supernovae (SN) within the Pantheon compilation and working within  flat $\Lambda$CDM cosmology, we observe a correlation between higher $H_0$ and the CMB dipole direction, confirming our previous results for strongly-lensed quasars \cite{Krishnan:2021dyb}. Concretely, we record a $\sim 1$ km/s/Mpc variation in $H_0$ at antipodal points on the sky within the Pantheon sample, which is evident in the Low $z$ subsample ($z \lesssim 0.075$) and gets enhanced by higher redshift SN. Our work raises the possibility that we may be at the precision required to probe anisotropic Hubble expansions, while providing a concrete prediction for future inferences of $H_0$.
\end{abstract}

\maketitle

\section{Introduction}
Systematics aside (most recently \cite{Efstathiou:2020wxn, Mortsell:2021nzg}), Hubble tension is a $\sim 10 \%$ discrepancy in the Hubble constant $H_0$ between an early Universe determination based on the flat $\Lambda$CDM model \cite{Aghanim:2018eyx} and late Universe determinations based on the distance ladder \cite{Riess:2019cxk, Riess:2020fzl}. To date, the search for cosmological solutions has been almost exclusively restricted to the Friedmann-Lema\^itre-Robertson-Walker (FLRW) paradigm  \cite{DiValentino:2021izs}. We recently laid bare the limitations of this approach by providing a ballpark figure for the maximum $H_0$ achievable within any FLRW cosmology \cite{Krishnan:2021dyb} (see also \cite{Vagnozzi:2021tjv}): 
\be
H_0 \sim 71 \pm \textrm{ 1 km/s/Mpc}, 
\ee
subject to the proviso that one works within General Relativity \footnote{Potential resolutions to both $H_0$ and $\sigma_8$ tensions may exist outside of GR, e. g. \cite{Sola:2021txs}.}. We stress that this  maximum $H_0$ can only be achieved if one allows an  alteration of the sound horizon with new early Universe physics. Evidently, local determinations of $H_0 \sim 73$ km/s/Mpc can be tolerated within $2 \sigma$, provided one ignores the tendency of new early Universe physics \cite{Poulin:2018cxd, Kreisch:2019yzn, AgRAwal:2019lmo, Niedermann:2019olb} to exacerbate $\sigma_8$ tension \cite{Hill:2020osr,Ivanov:2020ril,DAmico:2020ods, Niedermann:2020qbw, Murgia:2020ryi, Smith:2020rxx, Jedamzik:2020zmd, Lin:2021sfs}.

Over the last decade, we have witnessed a steady stream of independent studies ranging from radio galaxies \cite{Blake:2002gx, Singal:2011dy, Gibelyou:2012ri, Rubart:2013tx, Tiwari:2015tba, Colin:2017juj, Bengaly:2017slg, Siewert:2020krp} to quasars (QSOs) \cite{Secrest:2020has} showing that dipoles associated with CMB and distribution of large structures  do not match at large redshifts $z \gtrsim 1$ (even at $4.9 \sigma$ \cite{Secrest:2020has}) {(see also \cite{Singal:2021crs, Singal:2021kuu})}.
Throughout, the claims have been consistent that the dipole agrees with the CMB dipole in direction, but not magnitude. Once again, systematics could be at play, but it is worth noting that the redshift range is well beyond where peculiar velocities and local bulk flows are relevant \footnote{{Indeed, the methods used in these studies make no use of peculiar velocities and are based on distance-independent aberration and Doppler boosting analyses developed within the framework of special relativity that applies to distributions of sources at high redshifts \cite{ellisandbaldwin}.}}. Separately, it has been noted in galaxy cluster scaling relations \cite{Migkas:2020fza, Migkas:2021zdo} that $H_0$ may track the CMB dipole. Moreover, even with Planck CMB data, flat $\Lambda$CDM cosmological parameters exhibit dipoles close to the CMB dipole \cite{Yeung:2022smn}.
If true, this may point to a misinterpretation of the CMB dipole within modern cosmology. Simply put, observables should not track the CMB dipole when in CMB frame.

\begin{figure}[htb]
  \includegraphics[width=80mm]{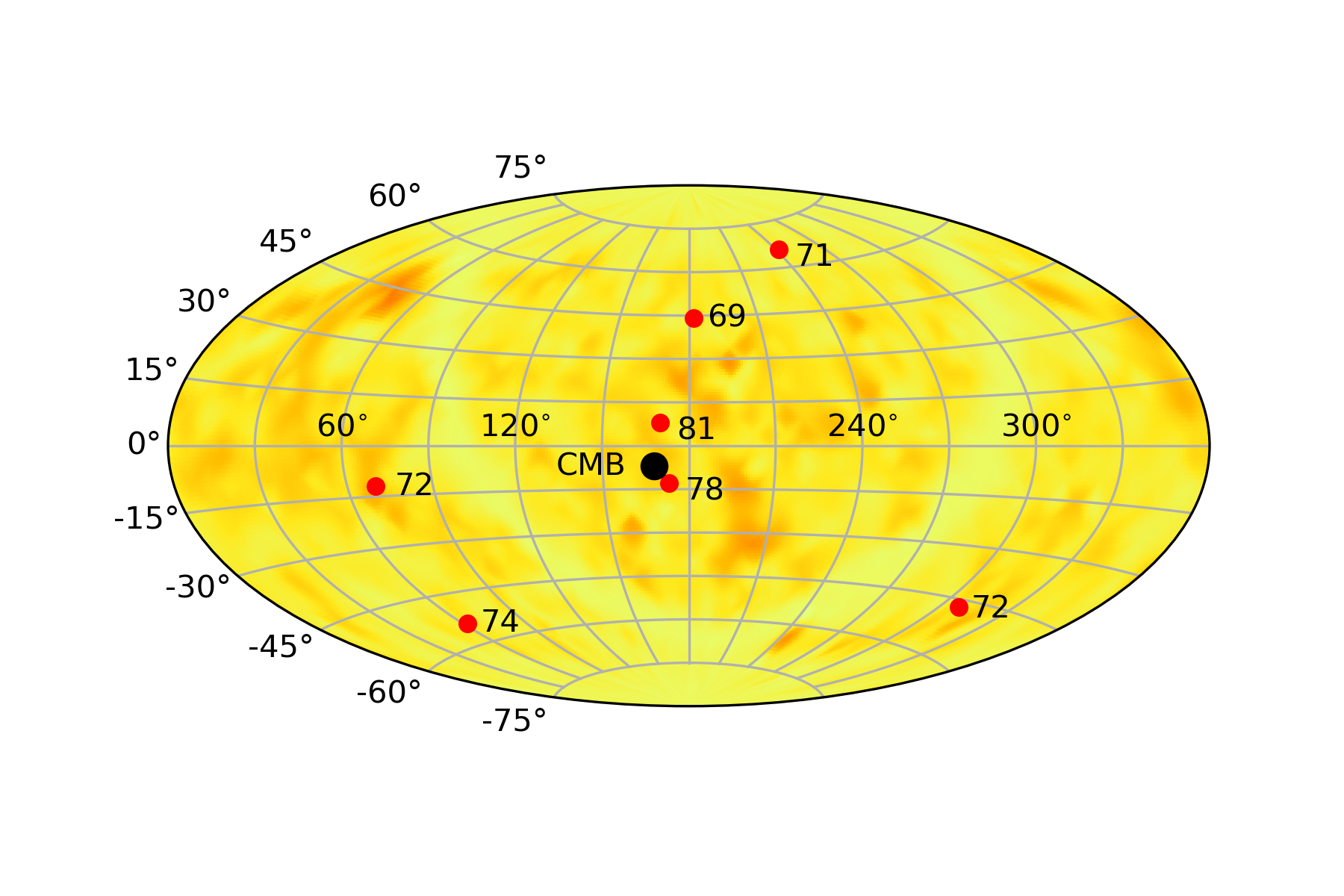}
\caption{$H_0$ values from strongly lensed QSOs alongside the CMB dipole against a background of 2MRS galaxy counts. These galaxies are relatively local and the observed anisotropies in 2MRS have to extend to  much higher redshifts for them to have a significant impact on the strong lensing measurements of the Hubble parameter.  
}
\label{lenses}
\end{figure}

Furthermore, it was recently observed \cite{Krishnan:2021dyb} that $H_0$ values inferred from strongly lensed QSOs within the flat $\Lambda$CDM model \cite{Wong:2019kwg, Millon:2019slk} appeared to be correlated with the CMB dipole direction. The same observation extends to $o$CDM and $w$CDM models \cite{Wong:2019kwg}. We illustrate this trend in Fig. \ref{lenses}, where we plot $H_0$ values against galaxy counts from the 2MRS catalog \cite{HuchRA:2011ii}, {where the latter gives an indication of structure in the local Universe.} 
There are three plausible explanations for Fig. \ref{lenses}:

\begin{enumerate}
\item This is a manifestation of a breakdown in FLRW at large redshift in line with persistent disagreement in the dipoles 
\cite{Blake:2002gx, Singal:2011dy, Gibelyou:2012ri, Rubart:2013tx, Tiwari:2015tba, Colin:2017juj, Bengaly:2017slg, Siewert:2020krp, Secrest:2020has}. 

\item Local structure along the line of sight could be playing a role \footnote{It may be a coincidence that the $H_0 \sim 74$ km/s/Mpc of DES lens \cite{2020MNRAS.494.6072S} is aligned with the Horologium-Reticulum supercluster. However, the 2MRS catalog is magnitude-limited at $K_s=11.75$ and therefore can barely reach the brightest galaxies beyond $c z \sim 15,000$ km/s and the clusters that make up Horologium-Reticulum appear to be closer to $c z \sim20,000$ km/s \cite{Fleenor:2005sf}.
\label{footnote-Horologium}}, underscoring a need to better model known superclusters when inferring $H_0$ from strong lensing time delay \cite{Wong:2019kwg, Millon:2019slk}. {Since the QSOs are deep in redshift, while the 2MRS catalog in Fig. \ref{lenses} is more local, one does not expect any immediate correlation.}

\item This could be a coincidence. 
\end{enumerate}

Already the balance may be tipped in favour of option 1, but here we support the observation further through Type Ia supernovae (SN). Recall that Cooke \& Lynden-Bell have already noted the presence of a small increase in the cosmic acceleration for the Union compilation of Type Ia SN \cite{Kowalski:2008ez} in the direction $( 198^{\circ},  -20^{\circ})$ \cite{Cooke:2009ws}. As emphasised in the original work \cite{Cooke:2009ws}, the effect is small $(\sim 1 \sigma)$, but it is apparent at higher redshifts, $z > 0.53$, {where it could overlap with any anisotropy traced by QSOs.} The main thrust of this short note is to revisit {and reinforce} this observation with the Pantheon data set \cite{Scolnic:2017caz}.

\section{Analysis}
{Before beginning, let us interpret Fig. \ref{lenses} as a statistical fluke in order to quantify how often this can happen. To do so, choose the CMB dipole direction, with right ascension ${\cal R}$ and declination ${\cal D}$, $({\cal R}, {\cal D})= (168^{\circ}, -7^{\circ})$. For each angle in Table IV of \cite{Krishnan:2021dyb}, construct the vector,
\be
\label{vec}
\vec{v} = \left( \cos {\cal D} \cos {\cal R} , \cos {\cal D} \sin {\cal R},  \sin {\cal D} \right). \\
\ee
Next, split the seven lenses into two hemispheres using the sign of the inner product of the CMB dipole direction vector with the lens vector. In each hemisphere $A$ \& $B$, one can identify a weighted average $H_0$ \footnote{{Since the errors on $H_0$ from the lenses are non-Gaussian we use the approximation $\sigma = \frac{1}{2} \left( 2 \sigma_1 \sigma_2/(\sigma_1 + \sigma_2) + \sqrt{\sigma_1 \sigma_2} \right)$.}}, before identifying the $H_0$ discrepancy:
\be
\label{sigma}
\sigma :=\frac{(H_0^{A} - H_0^{B})}{\sqrt{ (\delta H_0^A)^2 + (\delta H_0^B)^2}}. 
\ee
Concretely, we find $\sigma = 1.18$ for the real sample. To find out how representative this number is, we shift all lenses to their weighted averaged $H_0 = 73.53 \pm 1.46$ km/s/Mpc, and generate 100,000 copies of the original $H_0$ in a normal distribution using the $H_0$ error for each lens. Repeating the steps above, in 12\% of the configurations one finds a larger value of $\sigma$, so the probability of a fluke is $p = 0.12$. This may seem small compared to $\sigma = 1.18$, but note that $\sigma$ is a difference in weighted averages and the weighted averages have further internal freedoms. We have also repeated the exercise while weighting the $H_0$ values for both errors and orientation with respect to the CMB dipole, thereby bringing it more into line with later analysis. The difference was negligible, simply reducing $p = 0.12$ to $p = 0.116$.  
}

Next we move to Pantheon Type Ia SN, where we take all corrections for peculiar velocities at face value. The redshifts have been corrected for the local-group motion and are in the CMB rest frame ({\it e.g.} they are not heliocentric redshifts).  We refer the readers to \cite{Yang:2013gea, Javanmardi:2015sfa, Ghodsi:2016dwp, Andrade:2017iam, Bengaly:2018xko} for previous studies using SN to test the cosmological principle and to \cite{Deng:2018jrp, Chang:2019utc} for studies within Pantheon compilation and also to \cite{Colin:2010ds, Appleby:2014kea, Mohayaee:2020wxf} for the impact of peculiar velocities on nearby SN. In contrast to previous studies \cite{Deng:2018jrp, Chang:2019utc}, here we follow the intuition from Fig. \ref{lenses} and  \cite{Cooke:2009ws} that the CMB dipole direction is most relevant. {It should be stressed that it is \textit{assumed} the CMB dipole is simply due to relative motion, but an intrinsic dipole cannot yet be ruled out, (see {\it e.g.} \cite{FerreiRA:2020aqa}). Here, with limited assumptions we will show that a dipole emerges from the Pantheon data set, {\it despite the "sample" apparantly being cleaned for peculiar motions and being in the same rest frame as the CMB}. We first focus on the entire sample, which includes 1048 SN in the redshift range $0.01 < z \leq 2.26$ \footnote{We made use of the Open Supernova Catalog \cite{Guillochon:2016rhj} to identify 18 missing ${\cal R}, {\cal D}$ entries in the Pantheon data.}.} 

{Our analysis is similar to \cite{Schwarz:2007wf}, but we focus on $H_0$ rather than matter density $\Omega_{m0}$, since this is the most precisely constrained parameter by SN data, even at low redshifts, and it is the relevant parameter from Fig. \ref{lenses}. We work within the flat $\Lambda$CDM model and since $H_0$ and $\Omega_{m0}$ are negatively correlated, we have confirmed that scans for differences in $\Omega_{m0}$ produce similar results. As emphasised in \cite{Krishnan:2020vaf}, $H_0$ is an integration constant that is universal to \textit{all} FLRW cosmologies, whereas $\Omega_{m0}$ is a model dependent parameter. For this reason, we expect trends in $H_0$ to generalise to \textit{all} FLRW cosmologies, but obviously the significance will depend on the number of model parameters.} 

{We divide the sky up into a $31 \times 15$ grid for ${\cal R}$ and ${\cal D}$ in the ranges $0 \leq {\cal R} \leq 2 \pi$ and $-\pi/2 \leq {\cal D} \leq  \pi/2$. 
Note, there is some redundancy in the points at the boundaries of ${\cal R}$ and at the poles
\footnote{It is easy to increase the number of points in this grid and we expect that it will not greatly change our results.}. By opting for an odd number of points, we include the antipodal points on the sky, and as is clear from (\ref{vec}), by flipping the sign of ${\cal D}$  and shifting ${\cal R} \rightarrow {\cal R} \pm \pi$ one gets the antipodal point. For each point on the sky, we convert it to a vector $\vec{v}$ (\ref{vec}) and do the same }
for all the SN in the sample $\vec{v}_i$. Next, we split the SN according to the sign of the inner product $ \vec{v} \cdot \vec{v}_i$. SN with positive sign are in the hemisphere aligned with $\vec{v}$, whereas those with the negative sign are in the opposite hemisphere. One then fits the flat $\Lambda$CDM model in both hemispheres and identifies any discrepancy (\ref{sigma}), before repeating for all points on the sky. In performing the fitting, we make use of the full Pantheon covariance matrix \footnote{We find that removing the off-diagonal systematic uncertainties does not greatly affect our quoted significances}, i. e. both statistical and systematic uncertainties. {The latter takes the form of a covariance matrix \cite{Scolnic:2017caz}, which we crop accordingly to restrict to SN in a given hemisphere.} 

\begin{figure}[htb]
\centering
  \includegraphics[width=80mm]{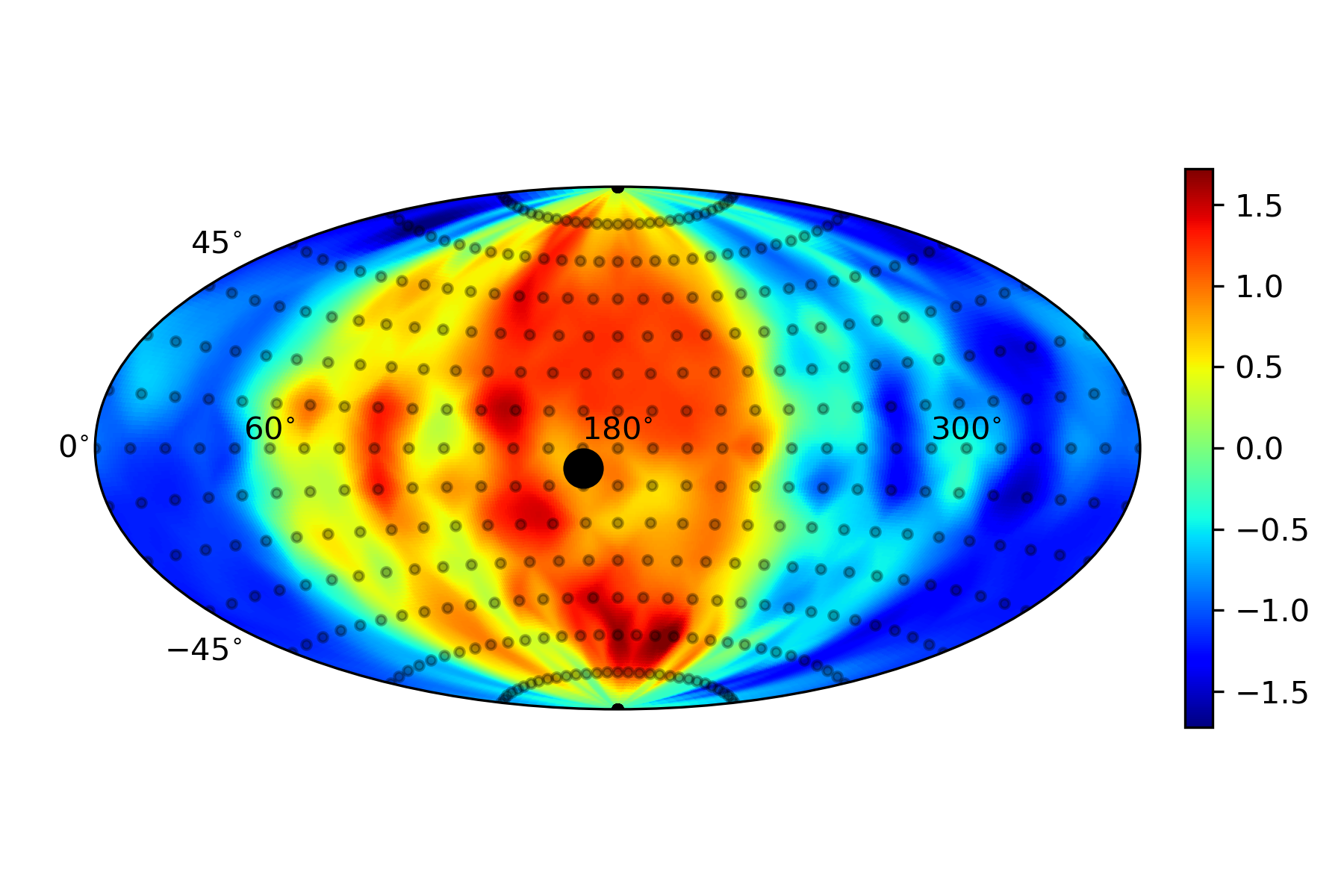}
\caption{Variation in the discrepancy in $H_0$ ($\sigma$) across the sky for the full Pantheon SN data set. The large black dot denotes the CMB dipole direction, while the smaller black dots show points on the sky where we have directly fitted flat $\Lambda$CDM prior to interpolation.
}
\label{scan1}
\end{figure}

{The result of the above exercise is shown in Fig. \ref{scan1}. Note that we have performed a cubic interpolation between points, using the python scipy library (scipy.interpolate.griddata) to improve presentation, but we only quote results for points where we have performed fitting.} As is clear from the plot, the higher values of $H_0$ (red regions) are in hemispheres aligned with the CMB dipole. The maximum value of $\sigma = 1.7$ is found at $ (216^{\circ}, -64.3^{\circ} )$. Thus, by symmetry, the minimum value of $\sigma = -1.7$ is at $(36^{\circ},+64.3^{\circ}$). In contrast, in the CMB dipole direction $\sigma$ is lower, $\sigma = 0.39$. However, it should be stressed that $H_0$ is more or less model agnostic within FLRW cosmologies and is one of the bluntest probes of anisotropy, since it has no directional dependence. Therefore, the main take away is the separation of the sky into higher $H_0$ and lower $H_0$ directions with the CMB dipole direction corresponding to a region of higher $H_0$. When phrased in absolute values, $\sim 1.5 \sigma$ equates to a $\Delta H_0 \sim 1$ km/s/Mpc ($\Delta m_{B} \sim 0.03$ mag) differential at antipodal points on the sky. This is comparable to the latest SH0ES $H_0$ errors \cite{Riess:2021jrx}. 

A number of comments are in order. First, we can view the most pronounced differences as a departure from FLRW, admittedly at low statistical significance. We can compare this intrinsic $\sim 1.5 \sigma$ dipole to a kinematic dipole by boosting the Pantheon redshifts to heliocentric frame following  \cite{Peterson:2021hel}. We have checked that this produces a stronger dipole at $\sim 4.5 \sigma$. For smaller kinematic velocities $v_0 < 369.82$ km/s, we find that the feature is rotated on the sky and cannot be removed by changing the magnitude of our relative velocity.
{Nevertheless, the trend in $H_0$ is consistent with Fig. \ref{lenses}.} Secondly, the orientation is consistent with the results of \cite{Blake:2002gx, Singal:2011dy, Gibelyou:2012ri, Rubart:2013tx, Tiwari:2015tba, Colin:2017juj, Bengaly:2017slg, Siewert:2020krp,  Secrest:2020has, Cai:2021wgv} in the sense that there is an unexpected dipole even though Pantheon SN are in the \textit{same frame} as the CMB. This effect, if physical, can potentially be traced to the difference in magnitude in the cosmic dipole reported in \cite{Blake:2002gx, Singal:2011dy, Gibelyou:2012ri, Rubart:2013tx, Tiwari:2015tba, Colin:2017juj, Bengaly:2017slg, Siewert:2020krp,  Secrest:2020has}. 

\begin{figure}[htb]
\centering
  \includegraphics[width=80mm]{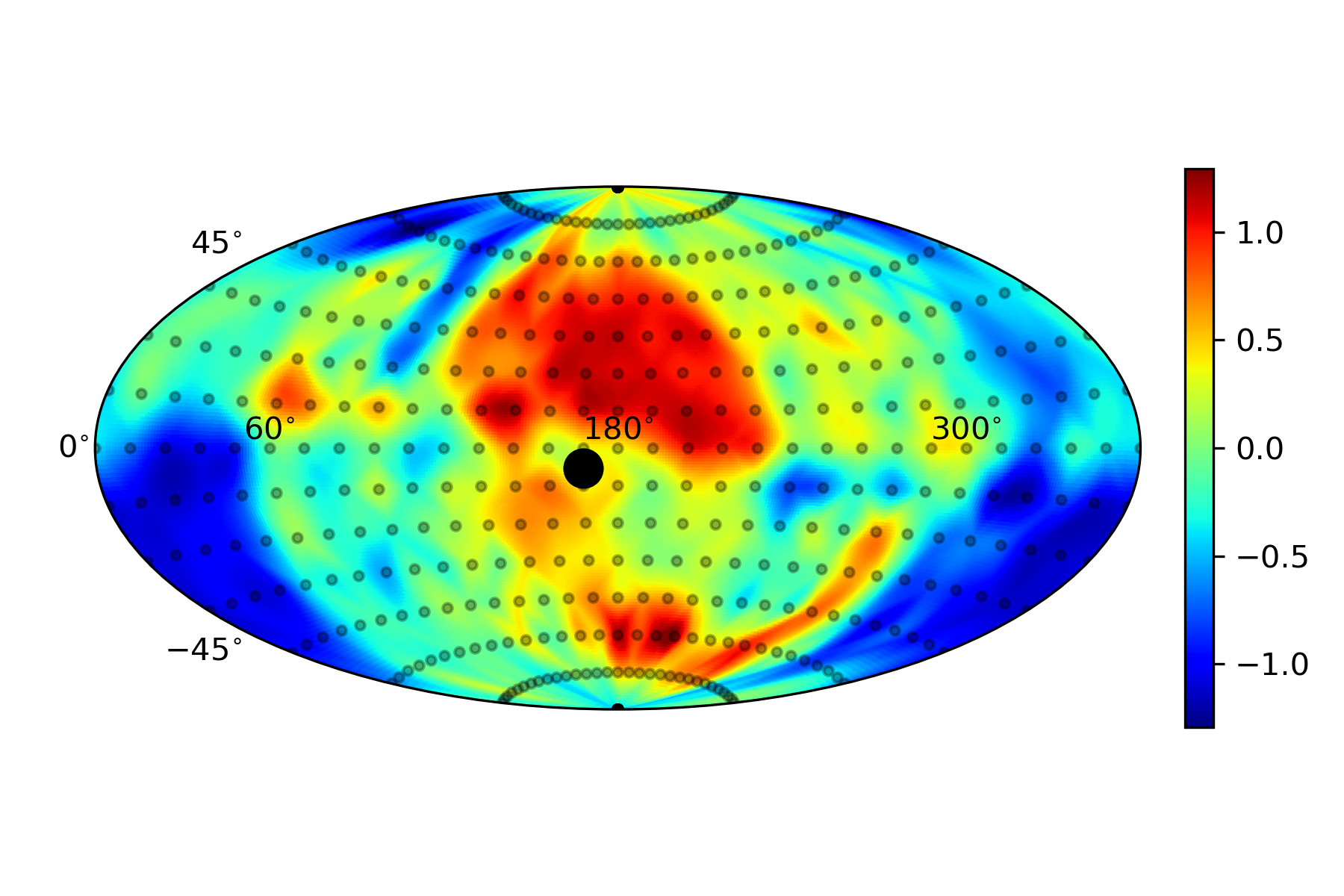}
\caption{Variation in the discrepancy in $H_0$ ($\sigma$) across the sky for the Low $z$ subsample ($0.01 < z \leq 0.07515$). The large black dot denotes the CMB dipole direction, while the smaller black dots show points on the sky where we have directly fitted flat $\Lambda$CDM prior to interpolation.}
\label{scan2}
\end{figure}

The Pantheon sample comprises Low $z$ \cite{Riess:1998dv, Jha:2005jg, Hicken:2009dk, Hicken:2009df, Hicken:2012zr, Contreras:2009nt, Folatelli:2009nm, Stritzinger:2011qd}  (172 SN), SDSS (335 SN) \cite{Frieman:2007mr, Kessler:2009ys, SDSS:2014irn},  PS1 (279 SN) \cite{Rest:2013mwz, Scolnic:2013efb}, SNLS (236 SN) \cite{SNLS:2011lii, SNLS:2011cra} and HST (26 SN) subsamples \cite{SupernovaCosmologyProject:2011ycw, SupernovaSearchTeam:2004lze, Riess:2006fw, Rodney:2014twa, Graur:2013msa, Riess:2017lxs}, with mean redshifts $\bar{z} = 0.03, 0.20, 0.29, 0.64$ and $1.27$, respectively (see \cite{Scolnic:2017caz} Fig. 10). However, neglecting the Low $z$ subsample, sky coverage is patchy: SDSS is exclusively in the hemisphere opposite to the CMB dipole, whereas PS1 covers 10 directions and SNLS samples 4 directions on the sky (3 of which are common to PS1). We have checked that only the removal of the Low $z$  subsample ($ 0.01 < z \leq 0.07515$) drastically affects Fig. \ref{scan1}. In short, while the other subsamples contribute to Fig. \ref{scan1}, the Low $z$ sample is central. Intuitively, this may have a simple explanation in the much better sky coverage. In Fig. \ref{scan2} we plot the same feature from the Low $z$ sample. Evidently, high redshift SN enhance the feature. It is worth noting that any residual dipole in the Low $z$ sample shows consistent directional dependence with reported anomalous bulk flows \cite{Watkins:2008hf, Lavaux:2008th, Colin:2010ds, 6dF}, which have most recently been recovered in Ref. \cite{Howlett:2022len}. Since we are working at low redshift, we have fixed $\Omega_{m0} = 0.3$. This restriction makes little difference as with fixed $\Omega_{m0}$, errors no longer propagate and differences in $H_0$ between hemispheres are still $\Delta H_0 \sim 1$ km/s/Mpc.

{Now comes the interesting question: how likely is Fig. \ref{scan1} as a statistical fluke? Before addressing this point, note that an \textit{a priori} exploration of the CMB dipole direction is well-motivated, since it represents the unique direction of motion of the local group with respect to the CMB. Moreover, a number of studies \cite{Cooke:2009ws, Siewert:2020krp, Secrest:2020has} have reported anisotropies in matter in directions consistent with the CMB dipole direction. If there is a dipole in matter that differs from the CMB dipole, it would be surprising if it exactly aligned, but close alignment may happen. Here, we use the CMB dipole direction within our local frame as an \textit{a priori} input in order to quantify the statistical significance of any departure from expect flat $\Lambda$CDM behaviour. }

{We essentially repeat the process leading to Fig. \ref{scan1}, but using a coarser sky grid $11 \times 5$. Once one accounts for boundaries, this leaves 50 independent, evenly spaced points on the sky. By symmetry, 25 of these points are in the same hemisphere as the CMB dipole direction. For each of these 25 points, we identify a $\sigma_{i}$ (\ref{sigma}), which we sum, but weighted according to the inner product of that direction on the sky with the CMB dipole direction, i. e.
\be
\sigma_{\textrm{sum}} = \sum_{i}  ( \vec{v}_i \cdot \vec{v}_{\textrm{dipole}}) \sigma_{i}.
\ee
This projects the 25 $\sigma_{i}$'s onto the CMB dipole direction, so that the points on the sky that are more closely aligned contribute more. The role of the weighted sum is to allow freedom in any putative matter dipole, while still focusing on directions close to the CMB dipole. }

\begin{figure}[htb]
\centering
  \includegraphics[width=80mm]{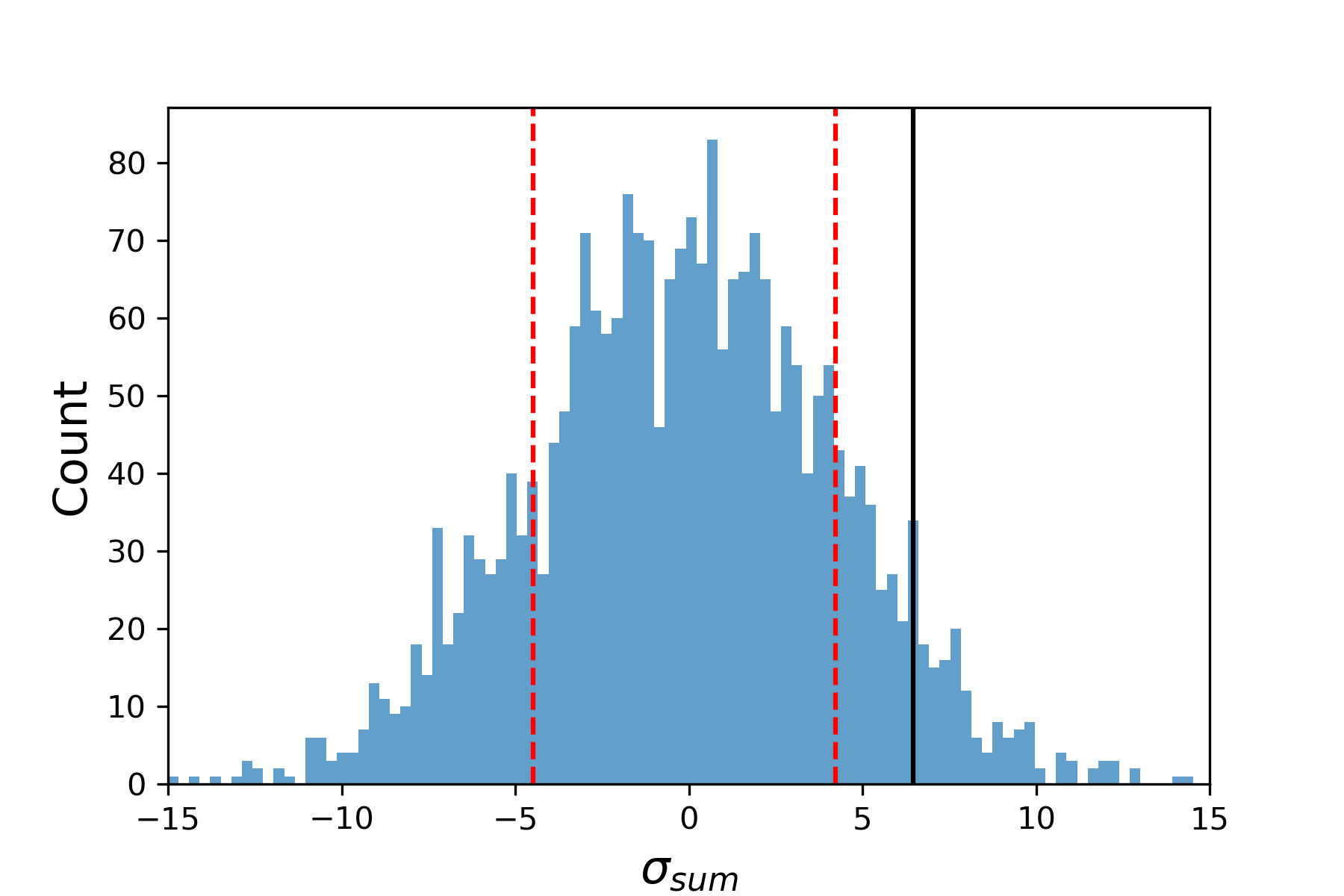}
\caption{The distribution of the weighted sum across 2500 mock realisations of the Pantheon data set. Red dotted lines denote $1 \sigma$, while the black line represents the value for the real data.}
\label{dist}
\end{figure}

{For the real data, we find $\sigma_{\textrm{sum}} = 6.471$, which one can compare to mock realisations of the data. To do so, we fit the flat $\Lambda$CDM model to the original data and extract best-fit values, $H_0 = 70.03 \pm 0.35$ km/s/Mpc, $\Omega_{m0} = 0.298 \pm 0.022$, and a corresponding $2 \times 2$ covariance matrix. Note that throughout} we fix the absolute magnitude $M$ of SN to a nominal constant value so that $H_0 \sim 70$ km/s/Mpc for the overall data set. Moreover, in fitting flat $\Lambda$CDM, we confirm the best-fit value of matter density from Table 8 of \cite{Scolnic:2017caz}. This serves as a rudimentary consistency check. {From the covariance matrix, we draw a normal distribution of 2500 $(H_0, \Omega_{m0})$ pairs, and for each pair, we generate values of apparent magnitude $m_{B}$ using the full covariance matrix of the Pantheon data set. For each realisation, we repeat the scan over the sky, identify $\sigma_{i}$ at 25 points, before identifying the corresponding weighted sum. We find a weighted sum exceeding the real data value in 162 from 2500 realisations, thus corresponding to a probability of $p = 0.065$. The resulting distribution of $\sigma_{\textrm{sum}}$ is shown in Fig. \ref{dist}.}  

\section{Outlook} 
As explained in \cite{Krishnan:2021dyb}, a resolution to Hubble tension within an FLRW cosmology requires the introduction of new early Universe physics. This is currently the front-running idea. However, it has become apparent that early Universe resolutions to Hubble tension typically exacerbate other discrepancies \cite{Hill:2020osr,Ivanov:2020ril,DAmico:2020ods, Niedermann:2020qbw, Murgia:2020ryi, Smith:2020rxx, Jedamzik:2020zmd, Lin:2021sfs}. Moreover, it should  be stressed that the introduction of new early Universe physics has only one real motivation, essentially to bring the Baryon Acoustic Oscillation (BAO) scale into line with Cepheid-calibrated SN, with no distinctive prediction. In physics, results lying on a single motivating observation, and without a predictive signature, are weak (even when they work). 

On the other hand, discrepancies between the CMB dipole and dipoles attributable to distant sources, namely radio galaxies \cite{Blake:2002gx, Singal:2011dy, Gibelyou:2012ri, Rubart:2013tx, Tiwari:2015tba, Colin:2017juj, Bengaly:2017slg, Siewert:2020krp} and QSOs \cite{Secrest:2020has}, have persisted for over a decade. These data are mostly at very high redshifts well beyond the redshifts where peculiar velocities/bulk flows can matter in FLRW.  

The correlation in Fig. \ref{lenses}, originally observed in  \cite{Krishnan:2021dyb}, may constitute a  fluke with probability $p=0.12$. Combining this with the probability we obtained here with Pantheon Type Ia SN \cite{Scolnic:2017caz}, using Fisher's method and noting that the strong lensing results (Fig.\ref{lenses}) and our present SN results (Fig.\ref{scan1}) are  statistically independent, yields $p = 0.046$, or alternatively, $1.7 \sigma$. The lensed QSO result, while low in statistical significance, is unlikely to be the result of some local effects, since the lenses and the QSOs are much deeper in redshift. Moreover, the consistency of SN and lensed QSO results with similar results in high redshift probes \cite{Luongo:2021nqh}, makes it  considerably less likely that our feature is a statistical fluke.

Certainly, Fig. \ref{scan1} is unexpected, since the Pantheon SN compilation has been put in CMB frame by construction, so any residual variation in $H_0$, especially one in the direction of the CMB dipole, is striking. While the various subsamples of Pantheon contribute to this feature, it is robust to the removal of all subsamples, except the Low $z$ subsample. Tellingly, this sample has the best sky coverage. Evidently, Fig. \ref{scan1} has other contributions beyond simply the Low $z$ subsample, but low redshift SN appear to play a key role (see also \cite{RAmeez:2019nrd, RAmeez:2019wdt, Mohayaee:2020wxf, Rubin:2019ywt, RAhman:2021mti}). Finally, it should be stressed that since we work within flat $\Lambda$CDM any
approach is minimal and a ``fundamental constant" within the FLRW paradigm, namely $H_0$, is under the spotlight.

Anisotropy in the Hubble parameter has previously been reported for the Union II dataset \cite{Cooke:2009ws}. Even prior to that and back in 2007 McClure and Dyer reported a statistically significant \textit{local} anisotropy in the value of Hubble parameter \cite{McClure:2007vv}. They used a single dataset from HST key project and are hence clean of systematics typical of composite datasets. That the Hubble parameter is locally anisotropic is not under question: local anisotropy in the Hubble value is usually absorbed into the peculiar velocities in perturbation analysis. The question is why such an anisotropy seems to persist in spite of Pantheon having undergone extensive cleaning for the local flow. The perhaps more fundamental question is why such an anisotropy seems to persist to high redshifts. This is totally unexpected in an FLRW Universe. 

In summary, there are a number of takeaways. Results from strong lensing data (Fig. \ref{lenses}) and SN data (Fig. \ref{scan1}) are consistent, thereby providing a concrete prediction for future $H_0$ probes ({\it e.g.} \cite{Luongo:2021nqh}). Secondly, our SN analysis (Fig. \ref{scan1}) shows both a low redshift and a high redshift effect, one which is not easily removed by changing the magnitude of our velocity with respect to CMB. Finally, $\Delta H_0 \sim 1$ km/s/Mpc is already at the level of precision of the latest SH0ES determination \cite{Riess:2021jrx}, which ultimately means that we may already be at a requisite precision to take anisotropic Hubble expansions seriously.

\section*{Acknowledgements} 

We thank Lucas Macri for correspondence. We thank Nikki Arendse, Angela Chen, Adri\`a Gomez-Valent, Jackson Said \& Joan Sol\`a for feedback on an earlier draft. {We are also grateful to anonymous referees for helping us sharpen our presentation.} E\'OC is funded by the National Research Foundation of Korea (NRF-2020R1A2C1102899).  MMShJ would like to acknowledge Saramadan grant No. ISEF/M/400122.
LY was supported by the National Research Foundation of Korea funded through the CQUeST of Sogang University (NRF-2020R1A6A1A03047877).

\appendix 

\section{Robustness of Observation}
\begin{figure}[htb]
\centering
  \includegraphics[width=80mm]{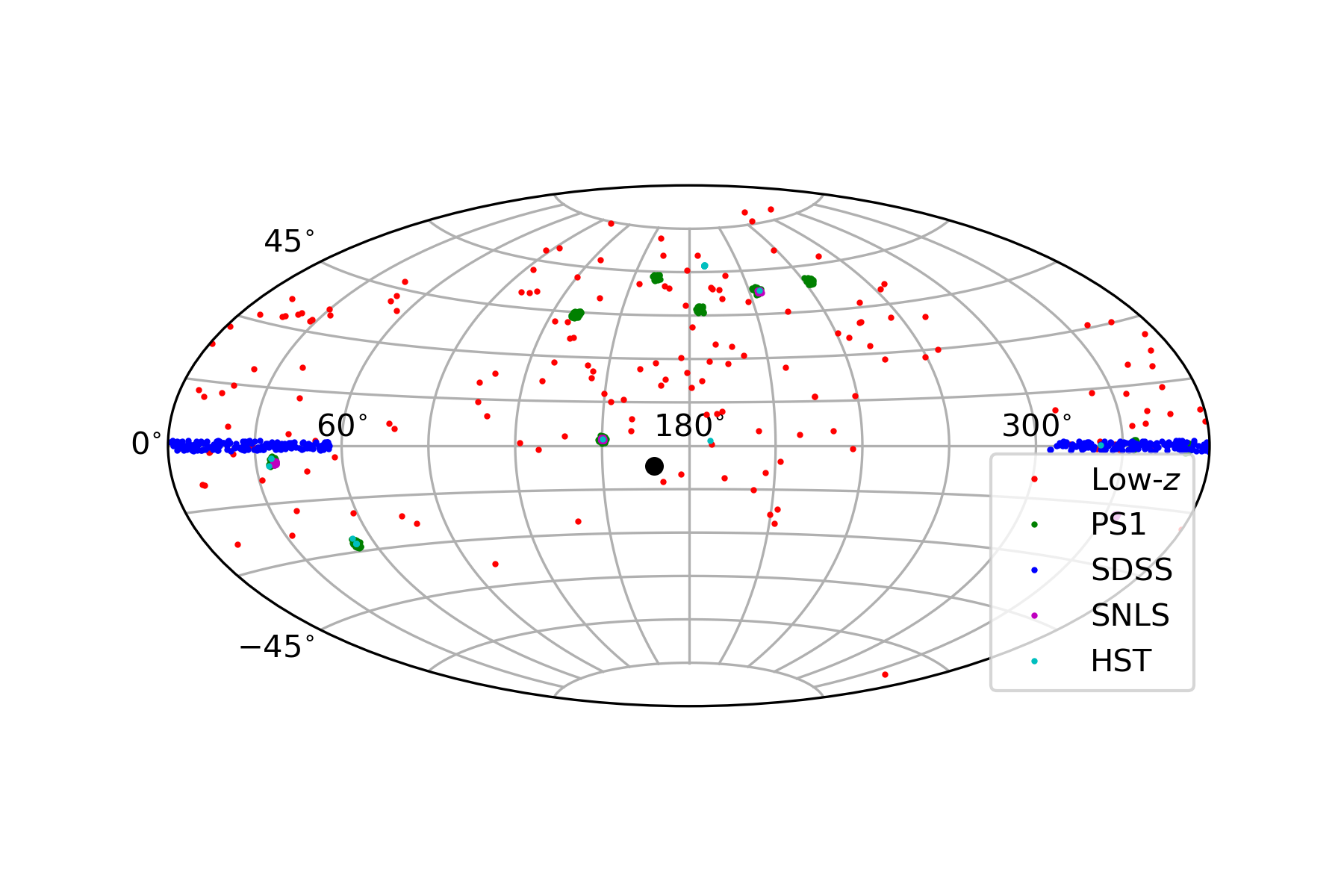}
\caption{Location of SN on the sky according to Pantheon subsample.}
\label{sne_loc}
\end{figure}

In this section, we provide some plots to show how the feature in Fig. \ref{scan1} changes as we remove SN subsamples. We begin with Fig. \ref{sne_loc} where we show the distribution of SN when decomposed into Low $z$ (172 SN), PS1 (279 SN), SDSS (335 SN), SNLS (236 SN) and HST SN (26 SN) subsamples. Understandably, as one moves to higher redshift, the sky coverage becomes pretty poor, but once seen, this is intuitive. The SDSS subsample (blue) is exclusively in the hemisphere opposite to the CMB dipole. Moreover, PS1 (green) covers only 10 (isolated) directions on the sky, while SNLS covers 4 (magenta) directions, of which 3 are common to PS1. In short, beyond Low $z$, the sky coverage is pretty poor. That being said, if the Universe is FLRW, this makes no difference. Observe also that there are relatively few SN in the lower hemisphere, whereas we are seeing a clear signal from that part of the sky in Fig. \ref{scan1}. The differences we report are between hemispheres, so they may be driven by SN in the opposite direction.

\begin{figure}[h]
\centering
  \includegraphics[width=80mm]{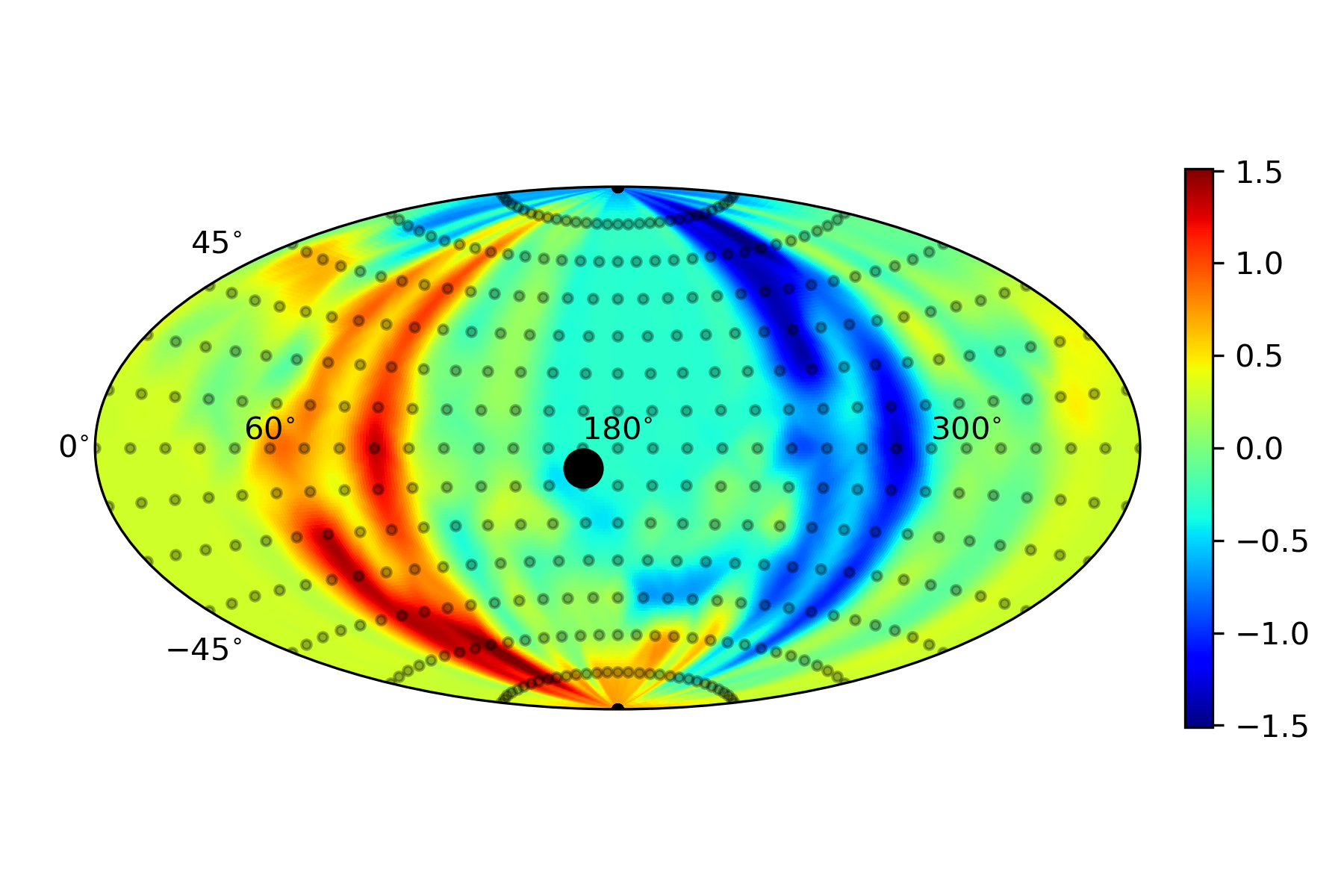}
\caption{Same as Fig. \ref{scan1} of manuscript but with Low $z$ removed.}
\label{no_lowz}
\end{figure}

\begin{figure}[h]
\centering
  \includegraphics[width=80mm]{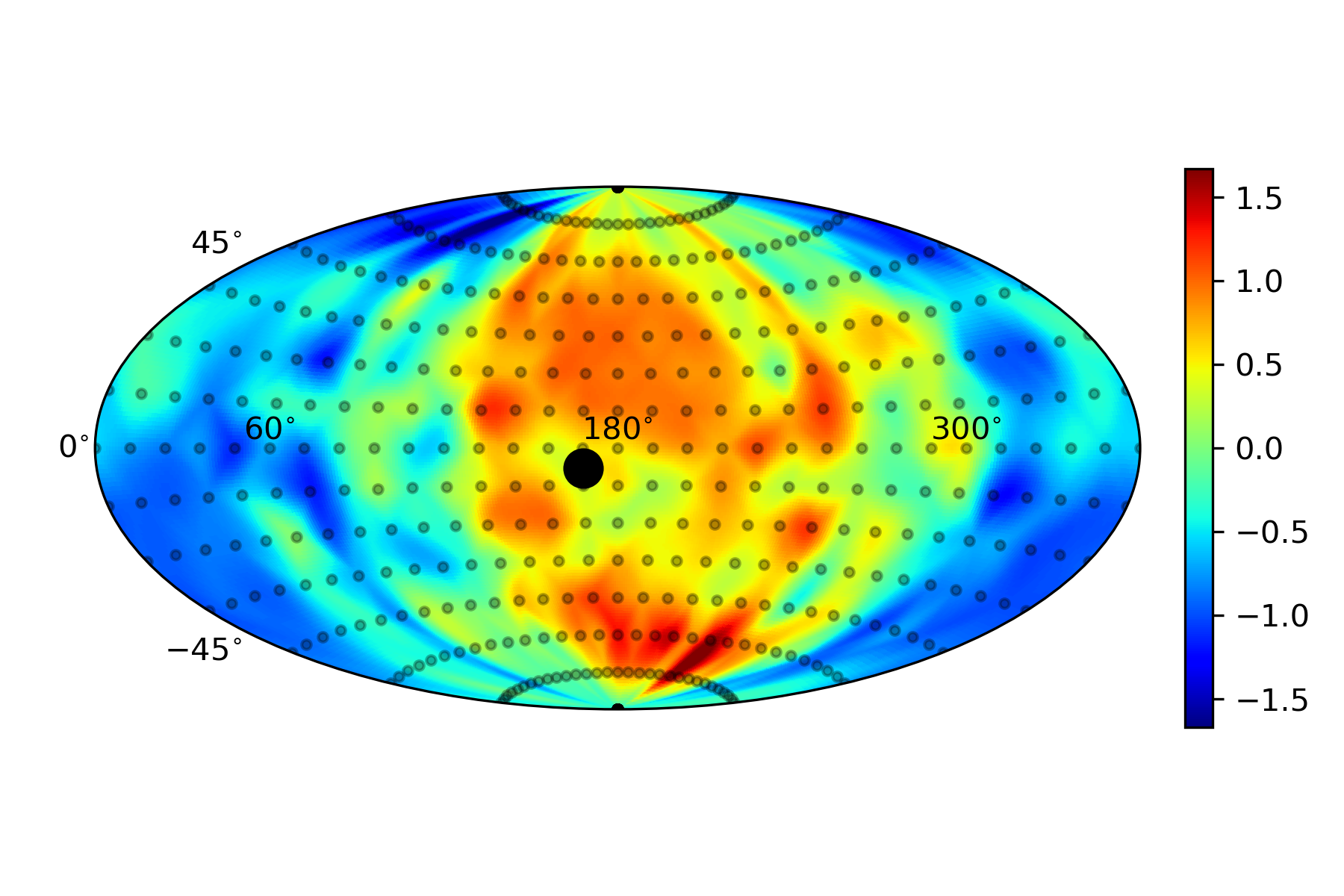}
\caption{Same as Fig. \ref{scan1} of manuscript but with SDSS removed.}
\label{no_SDSS}
\end{figure}

Next, we have repeated our scan by omitting each subsample in turn and, as expected, the most pronounced effect is found when the Low $z$ sample ($z \lesssim 0.075$) is removed. As is clear from Fig. \ref{no_lowz}, when the Low $z$ sample is removed, the dipole changes dramatically and one is no longer tracking the CMB dipole direction. Nevertheless, if one retains the Low $z$ sample, but remove the SDSS sample, which is the subsample with the next lowest effective redshift, one finds the feature in Fig. \ref{no_SDSS}. One can see that with Low $z$, but without SDSS, the plot starts to resemble Fig. \ref{scan1} in the sense that one is seeing variations in $H_0$ with greater significance, but still the feature is not so pronounced, yet evidently more pronounced than when only considers the Low $z$ sample in Fig. \ref{scan2}. Removing higher redshift samples beyond SDSS, namely PS1, SNLS or HST make very little difference, so we omit the plots. One interpretation of these results is that $H_0$ variations are mainly coming from Low $z$ sample, but the contributions from PS1, SDSS, SNLS and HST reinforce it. Let us stress that if there is a matter anisotropy aligned with the CMB dipole direction, then one will need SN samples with better overlap with these directions (see Fig. \ref{sne_loc}) to definitively confirm that higher redshift SN enhance the feature. 

\begin{figure}[h]
\centering
  \includegraphics[width=80mm]{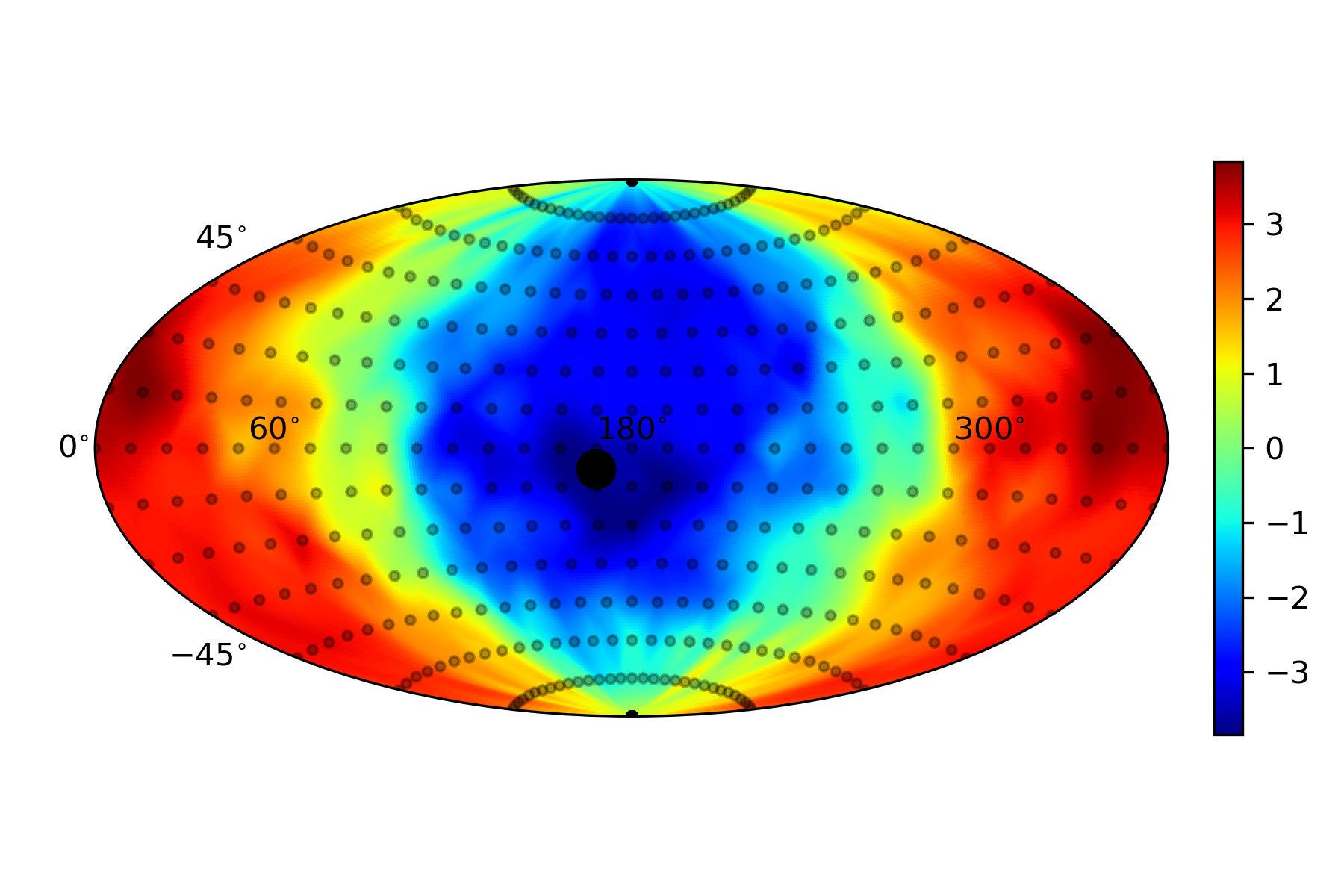}
\caption{We show the effect of boosting from $z_{CMB}$ to heliocentric frame. A kinematic dipole is considerably larger.}
\label{kinematic}
\end{figure}

Finally, we have performed some additional checks. Pantheon quotes both CMB redshifts $z_{CMB}$, some of which we have checked agree with the earlier JLA sample, but there are further corrected Hubble diagram redshifts, $z_{HD}$. The difference appears to be whether one corrects for any peculiar motion of the source SN, or not. These $z_{HD}$ are the redshifts that recover the cosmological parameters in the Pantheon paper \cite{Scolnic:2017caz}. We have checked that whether one uses $z_{CMB}$ or $z_{HD}$, Fig. \ref{scan1} in the manuscript is the same. One can also remove the systematic uncertainties from Pantheon, but this does not qualitatively change our Fig. 2. One still finds $\sim 1.5 \sigma$ discrepancies in certain directions on the sky. Nevertheless, boosting from CMB frame (starting from $z_{CMB}$) to heliocentric frame following equation (7) of Ref. \cite{Peterson:2021hel}, the significance increases by $\sim 4.5 \sigma$. To see this, note that we have an intrinsic dipole at $\sim 1.5 \sigma$ in Fig 2. of the draft, while in Fig. \ref{kinematic}, the colours are reversed at $\sim 3 \sigma$. This highlights the difference between the residual feature we see in Pantheon in \textit{CMB frame} (by construction) and a kinematic feature. A kinematic feature is expectedly stronger. So one of the take away messages of the manuscript is that despite all the corrections, Pantheon still has an intrinsic or residual dipole in the direction of the CMB dipole. This appears consistent with the lenses presented in Fig. \ref{lenses}.

\begin{thebibliography}{99}

\bibitem{Krishnan:2021dyb}
C.~Krishnan, R.~Mohayaee, E.~\'O~Colg\'ain, M.~M.~Sheikh-Jabbari and L.~Yin,
Class. Quant. GRAv. \textbf{38} (2021) no.18, 184001, [arXiv:2105.09790 [astro-ph.CO]].

\bibitem{Efstathiou:2020wxn}
G.~Efstathiou,
[arXiv:2007.10716 [astro-ph.CO]].

\bibitem{Mortsell:2021nzg}
E.~Mortsell, A.~Goobar, J.~Johansson and S.~Dhawan,
[arXiv:2105.11461 [astro-ph.CO]].

\bibitem{Aghanim:2018eyx} 
N.~Aghanim \textit{et al.} [Planck],
Astron. Astrophys. \textbf{641} (2020), A6
[erRAtum: Astron. Astrophys. \textbf{652} (2021), C4]
[arXiv:1807.06209 [astro-ph.CO]].


\bibitem{Riess:2019cxk}
A.~G.~Riess, S.~Casertano, W.~Yuan, L.~M.~Macri and D.~Scolnic,
Astrophys. J. \textbf{876} (2019) no.1, 85
[arXiv:1903.07603 [astro-ph.CO]].

\bibitem{Riess:2020fzl}
A.~G.~Riess, S.~Casertano, W.~Yuan, J.~B.~Bowers, L.~Macri, J.~C.~Zinn and D.~Scolnic,
Astrophys. J. Lett. \textbf{908} (2021) no.1, L6
[arXiv:2012.08534 [astro-ph.CO]].

\bibitem{DiValentino:2021izs}
E.~Di Valentino, O.~Mena, S.~Pan, L.~Visinelli, W.~Yang, A.~Melchiorri, D.~F.~Mota, A.~G.~Riess and J.~Silk,
Class. Quant. GRAv. \textbf{38} (2021) no.15, 153001
[arXiv:2103.01183 [astro-ph.CO]].

\bibitem{Vagnozzi:2021tjv}
S.~Vagnozzi, F.~Pacucci and A.~Loeb,
[arXiv:2105.10421 [astro-ph.CO]].

\bibitem{Sola:2021txs}
J.~Sol\`a Peracaula, A.~G\'omez-Valent, J.~de Cruz Perez and C.~Moreno-Pulido,
EPL \textbf{134} (2021) no.1, 19001
[arXiv:2102.12758 [astro-ph.CO]].

\bibitem{Poulin:2018cxd} 
  V.~Poulin, T.~L.~Smith, T.~Karwal and M.~Kamionkowski,
  Phys.\ Rev.\ Lett.\  {\bf 122}, no. 22, 221301 (2019)
  [arXiv:1811.04083 [astro-ph.CO]].

\bibitem{Kreisch:2019yzn} 
C.~D.~Kreisch, F.~Y.~Cyr-Racine and O.~Dor\'e,
Phys. Rev. D \textbf{101} (2020) no.12, 123505
[arXiv:1902.00534 [astro-ph.CO]].

 \bibitem{AgRAwal:2019lmo} 
  P.~Agrawal, F.~Y.~Cyr-Racine, D.~Pinner and L.~Randall,
  arXiv:1904.01016 [astro-ph.CO].

\bibitem{Niedermann:2019olb}
F.~Niedermann and M.~S.~Sloth,
Phys. Rev. D \textbf{103} (2021) no.4, L041303
[arXiv:1910.10739 [astro-ph.CO]]; 
Phys. Rev. D \textbf{102} (2020) no.6, 063527
[arXiv:2006.06686 [astro-ph.CO]].

\bibitem{Hill:2020osr}
J.~C.~Hill, E.~McDonough, M.~W.~Toomey and S.~Alexander,
Phys. Rev. D \textbf{102} (2020) no.4, 043507
[arXiv:2003.07355 [astro-ph.CO]].

\bibitem{Ivanov:2020ril}
M.~M.~Ivanov, E.~McDonough, J.~C.~Hill, M.~Simonovi\'c, M.~W.~Toomey, S.~Alexander and M.~Zaldarriaga,
Phys. Rev. D \textbf{102} (2020) no.10, 103502
[arXiv:2006.11235 [astro-ph.CO]].

\bibitem{DAmico:2020ods}
G.~D'Amico, L.~Senatore, P.~Zhang and H.~Zheng,
JCAP \textbf{05} (2021), 072
[arXiv:2006.12420 [astro-ph.CO]].

\bibitem{Niedermann:2020qbw}
F.~Niedermann and M.~S.~Sloth,
Phys. Rev. D \textbf{103} (2021) no.10, 103537
[arXiv:2009.00006 [astro-ph.CO]].

\bibitem{Murgia:2020ryi}
R.~Murgia, G.~F.~Abell\'an and V.~Poulin,
Phys. Rev. D \textbf{103} (2021) no.6, 063502
[arXiv:2009.10733 [astro-ph.CO]].

\bibitem{Smith:2020rxx}
T.~L.~Smith, V.~Poulin, J.~L.~Bernal, K.~K.~Boddy, M.~Kamionkowski and R.~Murgia,
Phys. Rev. D \textbf{103} (2021) no.12, 123542, [arXiv:2009.10740 [astro-ph.CO]].

\bibitem{Jedamzik:2020zmd}
K.~Jedamzik, L.~Pogosian and G.~B.~Zhao,
Commun. in Phys. \textbf{4} (2021), 123 
[arXiv:2010.04158 [astro-ph.CO]].

\bibitem{Lin:2021sfs}
W.~Lin, X.~Chen and K.~J.~Mack,
[arXiv:2102.05701 [astro-ph.CO]].

\bibitem{Blake:2002gx}
C.~Blake and J.~Wall,
Nature \textbf{416} (2002), 150-152
[arXiv:astro-ph/0203385 [astro-ph]].



\bibitem{Singal:2011dy}
A.~K.~Singal,
Astrophys. J. Lett. \textbf{742} (2011), L23
[arXiv:1110.6260 [astro-ph.CO]].

\bibitem{Gibelyou:2012ri}
C.~Gibelyou and D.~Huterer,
Mon. Not. Roy. Astron. Soc. \textbf{427} (2012), 1994-2021
[arXiv:1205.6476 [astro-ph.CO]].

\bibitem{Rubart:2013tx}
M.~Rubart and D.~J.~Schwarz,
Astron. Astrophys. \textbf{555} (2013), A117
[arXiv:1301.5559 [astro-ph.CO]].

\bibitem{Tiwari:2015tba}
P.~Tiwari and A.~Nusser,
JCAP \textbf{03} (2016), 062
[arXiv:1509.02532 [astro-ph.CO]].

\bibitem{Colin:2017juj}
J.~Colin, R.~Mohayaee, M.~Rameez and S.~Sarkar,
Mon. Not. Roy. Astron. Soc. \textbf{471} (2017) no.1, 1045-1055
[arXiv:1703.09376 [astro-ph.CO]].


\bibitem{Bengaly:2017slg}
C.~A.~P.~Bengaly, R.~Maartens and M.~G.~Santos,
JCAP \textbf{04} (2018), 031
[arXiv:1710.08804 [astro-ph.CO]].

\bibitem{Siewert:2020krp}
T.~M.~Siewert, M.~Schmidt-Rubart and D.~J.~Schwarz,
[arXiv:2010.08366 [astro-ph.CO]].

\bibitem{Secrest:2020has}
N.~J.~Secrest, S.~von Hausegger, M.~Rameez, R.~Mohayaee, S.~Sarkar and J.~Colin,
Astrophys. J. Lett. \textbf{908} (2021) no.2, L51
[arXiv:2009.14826 [astro-ph.CO]].
{
\bibitem{Cai:2021wgv}
R.~G.~Cai, Z.~K.~Guo, L.~Li, S.~J.~Wang and W.~W.~Yu,
Phys. Rev. D \textbf{103}, no.12, 121302 (2021)
[arXiv:2102.02020 [astro-ph.CO]].}

\bibitem{Singal:2021crs}
A.~K.~Singal,
[arXiv:2106.11968 [astro-ph.CO]].

\bibitem{Singal:2021kuu}
A.~K.~Singal,
[arXiv:2107.09390 [astro-ph.CO]].

\bibitem[\protect\citeauthoryear{Ellis \& Baldwin}{1984}]{ellisandbaldwin} Ellis G.~F.~R., Baldwin J.~E., 1984, MNRAS, 206, 377. 

\bibitem{Migkas:2020fza}
K.~Migkas, G.~Schellenberger, T.~H.~Reiprich, F.~Pacaud, M.~E.~Ramos-Ceja and L.~Lovisari,
Astron. Astrophys. \textbf{636} (2020), A15
[arXiv:2004.03305 [astro-ph.CO]].

\bibitem{Migkas:2021zdo}
K.~Migkas, F.~Pacaud, G.~Schellenberger, J.~Erler, N.~T.~Nguyen-Dang, T.~H.~Reiprich, M.~E.~Ramos-Ceja and L.~Lovisari,
Astron. Astrophys. \textbf{649} (2021), A151
[arXiv:2103.13904 [astro-ph.CO]].

\bibitem{Yeung:2022smn}
S.~Yeung and M.~C.~Chu,
``Directional Variations of Cosmological Parameters from the 
[arXiv:2201.03799 [astro-ph.CO]].

\bibitem{Wong:2019kwg}
K.~C.~Wong, S.~H.~Suyu, \textit{et al.} (H0LiCOW collaboration)
Mon. Not. Roy. Astron. Soc. \textbf{498} (2020) no.1, 1420-1439
[arXiv:1907.04869 [astro-ph.CO]].



\bibitem{Millon:2019slk}
M.~Millon, A.~Galan, F.~Courbin, T.~Treu, S.~H.~Suyu, X.~Ding, S.~Birrer, G.~C.~F.~Chen, A.~J.~Shajib and D.~Sluse, \textit{et al.}
Astron. Astrophys. \textbf{639} (2020), A101
[arXiv:1912.08027 [astro-ph.CO]].

\bibitem{HuchRA:2011ii}
J.~P.~Huchra, L.~M.~Macri, K.~L.~Masters, T.~H.~Jarrett, P.~Berlind, M.~Calkins, A.~C.~Crook, R.~Cutri, P.~Erdogdu and E.~Falco, \textit{et al.}
Astrophys. J. Suppl. \textbf{199} (2012), 26
[arXiv:1108.0669 [astro-ph.CO]].

\bibitem{2020MNRAS.494.6072S} 
A.~J.~Shajib \textit{et al.} [DES],
Mon. Not. Roy. Astron. Soc. \textbf{494} (2020) no.4, 6072-6102
[arXiv:1910.06306 [astro-ph.CO]].

\bibitem{Fleenor:2005sf}
M.~C.~Fleenor, J.~A.~Rose, W.~A.~Christiansen, M.~Johnston-Hollitt, R.~W.~Hunstead, M.~J.~Drinkwater and W.~Saunders,
Astron. J. \textbf{131} (2006), 1280-1287
[arXiv:astro-ph/0512169 [astro-ph]].

\bibitem{Kowalski:2008ez}
M.~Kowalski \textit{et al.} [Supernova Cosmology Project],
Astrophys. J. \textbf{686} (2008), 749-778
[arXiv:0804.4142 [astro-ph]].

\bibitem{Cooke:2009ws}
R.~Cooke and D.~Lynden-Bell,
Mon. Not. Roy. Astron. Soc. \textbf{401} (2010), 1409-1414
[arXiv:0909.3861 [astro-ph.CO]].

\bibitem{Scolnic:2017caz}
D.~Scolnic \textit{et al.},
Astrophys. J. \textbf{859} (2018) no.2, 101
[arXiv:1710.00845 [astro-ph.CO]].

\bibitem{RAmeez:2019nrd}
M.~Rameez,
[arXiv:1905.00221 [astro-ph.CO]].

\bibitem{RAmeez:2019wdt}
M.~Rameez and S.~Sarkar,
Class. Quant. Grav. \textbf{38} (2021) no.15, 154005 [arXiv:1911.06456 [astro-ph.CO]].

\bibitem{Mohayaee:2020wxf}
R.~Mohayaee, M.~Rameez and S.~Sarkar,
[arXiv:2003.10420 [astro-ph.CO]].

\bibitem{Rubin:2019ywt}
D.~Rubin and J.~Heitlauf,
Astrophys. J. \textbf{894} (2020) no.1, 68
[arXiv:1912.02191 [astro-ph.CO]].

\bibitem{RAhman:2021mti}
W.~Rahman, R.~Trotta, S.~S.~Boruah, M.~J.~Hudson and D.~A.~van Dyk,
[arXiv:2108.12497 [astro-ph.CO]].

\bibitem{Colgain:2019pck}
E.~\'O~Colg\'ain,
JCAP \textbf{09} (2019), 006
[arXiv:1903.11743 [astro-ph.CO]].

\bibitem{mvp2020}
Maurice H P M van Putten, 
MNRAS Volume 491, Issue 1, January 2020, Pages L6-L10

\bibitem{Javanmardi:2015sfa}
B.~Javanmardi, C.~Porciani, P.~Kroupa and J.~Pflamm-Altenburg,
Astrophys. J. \textbf{810} (2015) no.1, 47
[arXiv:1507.07560 [astro-ph.CO]].

\bibitem{Ghodsi:2016dwp}
H.~Ghodsi, S.~Baghram and F.~Habibi,
JCAP \textbf{10} (2017), 017
[arXiv:1609.08012 [astro-ph.CO]].

\bibitem{Bengaly:2018xko}
C.~A.~P.~Bengaly, U.~Andrade and J.~S.~Alcaniz,
Eur. Phys. J. C \textbf{79} (2019) no.9, 768
[arXiv:1810.04966 [astro-ph.CO]].

\bibitem{Andrade:2017iam}
U.~Andrade, C.~A.~P.~Bengaly, J.~S.~Alcaniz and B.~Santos,
Phys. Rev. D \textbf{97} (2018) no.8, 083518
[arXiv:1711.10536 [astro-ph.CO]].

\bibitem{Yang:2013gea}
X.~Yang, F.~Y.~Wang and Z.~Chu,
Mon. Not. Roy. Astron. Soc. \textbf{437} (2014) no.2, 1840-1846
[arXiv:1310.5211 [astro-ph.CO]].

\bibitem{Deng:2018jrp}
H.~K.~Deng and H.~Wei,
Eur. Phys. J. C \textbf{78} (2018) no.9, 755
[arXiv:1806.02773 [astro-ph.CO]].

\bibitem{Chang:2019utc}
Z.~Chang, D.~Zhao and Y.~Zhou,
Chin. Phys. C \textbf{43} (2019) no.12, 125102
[arXiv:1910.06883 [astro-ph.CO]].

\bibitem{Guillochon:2016rhj}
J.~Guillochon, J.~Parrent, L.~Z.~Kelley and R.~Margutti,
Astrophys. J. \textbf{835} (2017) no.1, 64
[arXiv:1605.01054 [astro-ph.SR]].

\bibitem{Colin:2010ds}
J.~Colin, R.~Mohayaee, S.~Sarkar and A.~Shafieloo,
Mon. Not. Roy. Astron. Soc. \textbf{414} (2011), 264-271
[arXiv:1011.6292 [astro-ph.CO]].

\bibitem{Appleby:2014kea}
S.~Appleby, A.~Shafieloo and A.~Johnson,
Astrophys. J. \textbf{801} (2015) no.2, 76
[arXiv:1410.5562 [astro-ph.CO]].

\bibitem{FerreiRA:2020aqa}
P.~d.~Ferreira and M.~Quartin,
Phys. Rev. Lett. \textbf{127} (2021) no.10, 101301
[arXiv:2011.08385 [astro-ph.CO]].

\bibitem{Schwarz:2007wf}
D.~J.~Schwarz and B.~Weinhorst,
Astron. Astrophys. \textbf{474} (2007), 717-729
[arXiv:0706.0165 [astro-ph]].

\bibitem{Krishnan:2020vaf}
C.~Krishnan, E.~\'O~Colg\'ain, M.~M.~Sheikh-Jabbari and T.~Yang,
Phys. Rev. D \textbf{103} (2021) no.10, 103509
[arXiv:2011.02858 [astro-ph.CO]].

\bibitem{Riess:2021jrx}
A.~G.~Riess, W.~Yuan, L.~M.~Macri, D.~Scolnic, D.~Brout, S.~Casertano, D.~O.~Jones, Y.~Murakami, L.~Breuval and T.~G.~Brink, \textit{et al.}
[arXiv:2112.04510 [astro-ph.CO]].

\bibitem{Peterson:2021hel}
E.~R.~Peterson, W.~D'Arcy Kenworthy, D.~Scolnic, A.~G.~Riess, D.~Brout, A.~Carr, H.~Courtois, T.~Davis, A.~Dwomoh and D.~O.~Jones, \textit{et al.}
[arXiv:2110.03487 [astro-ph.CO]].

\bibitem{Riess:1998dv}
A.~G.~Riess, R.~P.~Kirshner, B.~P.~Schmidt, S.~Jha, P.~Challis, P.~M.~Garnavich, A.~A.~Esin, C.~Carpenter, R.~Grashius and R.~E.~Schild, \textit{et al.}
Astron. J. \textbf{117} (1999), 707-724
[arXiv:astro-ph/9810291 [astro-ph]].

\bibitem{Jha:2005jg}
S.~Jha, R.~P.~Kirshner, P.~Challis, P.~M.~Garnavich, T.~Matheson, A.~M.~Soderberg, G.~J.~M.~Graves, M.~Hicken, J.~F.~Alves and H.~G.~Arce, \textit{et al.}
Astron. J. \textbf{131} (2006), 527-554
[arXiv:astro-ph/0509234 [astro-ph]].

\bibitem{Hicken:2009dk}
M.~Hicken, W.~M.~Wood-Vasey, S.~Blondin, P.~Challis, S.~Jha, P.~L.~Kelly, A.~Rest and R.~P.~Kirshner,
Astrophys. J. \textbf{700} (2009), 1097-1140
[arXiv:0901.4804 [astro-ph.CO]].

\bibitem{Hicken:2009df}
M.~Hicken, P.~Challis, S.~Jha, R.~P.~Kirsher, T.~Matheson, M.~Modjaz, A.~Rest and W.~M.~Wood-Vasey,
Astrophys. J. \textbf{700} (2009), 331-357
[arXiv:0901.4787 [astro-ph.CO]].

\bibitem{Hicken:2012zr}
M.~Hicken, P.~Challis, R.~P.~Kirshner, A.~Rest, C.~E.~Cramer, W.~M.~Wood-Vasey, G.~Bakos, P.~Berlind, W.~R.~Brown and N.~Caldwell, \textit{et al.}
Astrophys. J. Suppl. \textbf{200} (2012), 12
[arXiv:1205.4493 [astro-ph.CO]].

\bibitem{Contreras:2009nt}
C.~Contreras, M.~Hamuy, M.~M.~Phillips, G.~Folatelli, N.~B.~Suntzeff, S.~E.~Persson, M.~Stritzinger, L.~Boldt, S.~Gonzalez and W.~Krzeminski, \textit{et al.}
Astron. J. \textbf{139} (2010), 519-539
[arXiv:0910.3330 [astro-ph.CO]].

\bibitem{Folatelli:2009nm}
G.~Folatelli, M.~M.~Phillips, C.~R.~Burns, C.~Contreras, M.~Hamuy, W.~L.~Freedman, S.~E.~Persson, M.~Stritzinger, N.~B.~Suntzeff and K.~Krisciunas, \textit{et al.}
Astron. J. \textbf{139} (2010), 120-144
[arXiv:0910.3317 [astro-ph.CO]].

\bibitem{Stritzinger:2011qd}
M.~D.~Stritzinger, M.~M.~Phillips, L.~N.~Boldt, C.~Burns, A.~Campillay, C.~Contreras, S.~Gonzalez, G.~Folatelli, N.~Morrell and W.~Krzeminski, \textit{et al.}
Astron. J. \textbf{142} (2011), 156
[arXiv:1108.3108 [astro-ph.CO]].

\bibitem{Frieman:2007mr}
J.~A.~Frieman, B.~Bassett, A.~Becker, C.~Choi, D.~Cinabro, D.~F.~DeJongh, D.~L.~Depoy, M.~Doi, P.~M.~Garnavich and C.~J.~Hogan, \textit{et al.}
Astron. J. \textbf{135} (2008), 338-347
[arXiv:0708.2749 [astro-ph]].

\bibitem{Kessler:2009ys}
R.~Kessler, A.~Becker, D.~Cinabro, J.~Vanderplas, J.~A.~Frieman, J.~Marriner, T.~M.~Davis, B.~Dilday, J.~Holtzman and S.~Jha, \textit{et al.}
Astrophys. J. Suppl. \textbf{185} (2009), 32-84
[arXiv:0908.4274 [astro-ph.CO]].

\bibitem{SDSS:2014irn}
M.~Sako \textit{et al.} [SDSS],
Publ. Astron. Soc. Pac. \textbf{130} (2018) no.988, 064002
[arXiv:1401.3317 [astro-ph.CO]].

\bibitem{Rest:2013mwz}
A.~Rest, D.~Scolnic, R.~J.~Foley, M.~E.~Huber, R.~Chornock, G.~Narayan, J.~L.~Tonry, E.~Berger, A.~M.~Soderberg and C.~W.~Stubbs, \textit{et al.}
Astrophys. J. \textbf{795} (2014) no.1, 44
[arXiv:1310.3828 [astro-ph.CO]].

\bibitem{Scolnic:2013efb}
D.~Scolnic, A.~Rest, A.~Riess, M.~E.~Huber, R.~J.~Foley, D.~Brout, R.~Chornock, G.~Narayan, J.~L.~Tonry and E.~Berger, \textit{et al.}
Astrophys. J. \textbf{795} (2014) no.1, 45
[arXiv:1310.3824 [astro-ph.CO]].

\bibitem{SNLS:2011lii}
A.~Conley \textit{et al.} [SNLS],
Astrophys. J. Suppl. \textbf{192} (2011), 1
[arXiv:1104.1443 [astro-ph.CO]].

\bibitem{SNLS:2011cra}
M.~Sullivan \textit{et al.} [SNLS],
Astrophys. J. \textbf{737} (2011), 102
[arXiv:1104.1444 [astro-ph.CO]].

\bibitem{SupernovaCosmologyProject:2011ycw}
N.~Suzuki \textit{et al.} [Supernova Cosmology Project],
Astrophys. J. \textbf{746} (2012), 85
[arXiv:1105.3470 [astro-ph.CO]].

\bibitem{SupernovaSearchTeam:2004lze}
A.~G.~Riess \textit{et al.} [Supernova Search Team],
Astrophys. J. \textbf{607} (2004), 665-687
[arXiv:astro-ph/0402512 [astro-ph]].

\bibitem{Riess:2006fw}
A.~G.~Riess, L.~G.~Strolger, S.~Casertano, H.~C.~Ferguson, B.~Mobasher, B.~Gold, P.~J.~Challis, A.~V.~Filippenko, S.~Jha and W.~Li, \textit{et al.}
Astrophys. J. \textbf{659} (2007), 98-121
[arXiv:astro-ph/0611572 [astro-ph]].

\bibitem{Rodney:2014twa}
S.~A.~Rodney, A.~G.~Riess, L.~G.~Strolger, T.~Dahlen, O.~Graur, S.~Casertano, M.~E.~Dickinson, H.~C.~Ferguson, P.~Garnavich and B.~Hayden, \textit{et al.}
Astron. J. \textbf{148} (2014), 13
[arXiv:1401.7978 [astro-ph.CO]].

\bibitem{Graur:2013msa}
O.~Graur, S.~A.~Rodney, D.~Maoz, A.~G.~Riess, S.~W.~Jha, M.~Postman, T.~Dahlen, T.~W.~S.~Holoien, C.~McCully and B.~Patel, \textit{et al.}
Astrophys. J. \textbf{783} (2014), 28
[arXiv:1310.3495 [astro-ph.CO]].

\bibitem{Riess:2017lxs}
A.~G.~Riess, S.~A.~Rodney, D.~M.~Scolnic, D.~L.~Shafer, L.~G.~Strolger, H.~C.~Ferguson, M.~Postman, O.~Graur, D.~Maoz and S.~W.~Jha, \textit{et al.}
Astrophys. J. \textbf{853} (2018) no.2, 126
[arXiv:1710.00844 [astro-ph.CO]].

\bibitem{Lusso:2020pdb}
E.~Lusso, G.~Risaliti, E.~Nardini, G.~Bargiacchi, M.~Benetti, S.~Bisogni, S.~Capozziello, F.~Civano, L.~Eggleston and M.~Elvis, \textit{et al.}
Astron. Astrophys. \textbf{642} (2020), A150
[arXiv:2008.08586 [astro-ph.GA]].

\bibitem{Demianski:2016zxi}
M.~Demianski, E.~Piedipalumbo, D.~Sawant and L.~Amati,
Astron. Astrophys. \textbf{598} (2017), A112
[arXiv:1610.00854 [astro-ph.CO]].

\bibitem{Watkins:2008hf}
R.~Watkins, H.~A.~Feldman and M.~J.~Hudson,
Mon. Not. Roy. Astron. Soc. \textbf{392} (2009), 743-756
[arXiv:0809.4041 [astro-ph]].

\bibitem{Lavaux:2008th}
G.~Lavaux, R.~B.~Tully, R.~Mohayaee and S.~Colombi,
Astrophys. J. \textbf{709} (2010), 483-498
[arXiv:0810.3658 [astro-ph]].

\bibitem{6dF}
Magoulas, C., Springob, C., Colless, M., Mould, J., Lucey, J., Erdogdu, P., \& Jones, D. (2014). Proceedings of the International Astronomical Union, 11(S308), 336-339. 

\bibitem{Howlett:2022len}
C.~Howlett, K.~Said, J.~R.~Lucey, M.~Colless, F.~Qin, Y.~Lai, R.~B.~Tully and T.~M.~Davis,
[arXiv:2201.03112 [astro-ph.CO]].

\bibitem{Luongo:2021nqh}
O.~Luongo, M.~Muccino, E.~\'O~Colg\'ain, M.~M.~Sheikh-Jabbari and L.~Yin,
[arXiv:2108.13228 [astro-ph.CO]].

\bibitem{McClure:2007vv}
M.~L.~McClure and C.~C.~Dyer,
New Astron. \textbf{12} (2007), 533-543
[arXiv:astro-ph/0703556 [astro-ph]].

\bibitem{Peterson:2021hel}
E.~R.~Peterson, W.~D'Arcy Kenworthy, D.~Scolnic, A.~G.~Riess, D.~Brout, A.~Carr, H.~Courtois, T.~Davis, A.~Dwomoh and D.~O.~Jones, \textit{et al.}
[arXiv:2110.03487 [astro-ph.CO]].




\end{thebibliography}
\end{document}